\documentclass{aa}

\usepackage{graphicx}
\usepackage[varg]{txfonts}

\newcommand{\msol}{\mbox{${\rm M}_{\sun}$}}

\begin{document}
\title{Deriving physical parameters of unresolved star clusters.}
\subtitle{I. Age, mass, and extinction degeneracies}

\author{P. de Meulenaer\inst{1,2} \and D. Narbutis\inst{1} \and T. Mineikis\inst{1,2} \and V. Vansevi\v{c}ius\inst{1,2}}

\institute{Vilnius University Observatory, \v{C}iurlionio 29, Vilnius LT-03100, Lithuania\\\email{philippe.demeulenaer@ff.stud.vu.lt} \and Center for Physical Sciences and Technology, Savanori\c{u} 231, Vilnius LT-02300, Lithuania}

\date{Submitted October 31, 2012; Accepted November 28, 2012}

\abstract
{Stochasticity and physical parameter degeneracy problems complicate the derivation of the parameters (age, mass, and extinction) of unresolved star clusters when using broad-band photometry.}
{We develop a method to simulate stochasticity and degeneracies, and to investigate their influence on the accuracy of derived physical parameters. Then we apply it to star cluster samples of M31 and M33 galaxies.}
{Age, mass and extinction of observed star clusters are derived by comparing their broad-band $UBVRI$ integrated magnitudes to the magnitudes of a large grid of star cluster models with fixed metallicity $Z=0.008$. Masses of stars for a cluster model are randomly sampled from the initial mass function. Models of star clusters from the model grid, which have all of their magnitudes located within 3 observational errors from the magnitudes of the observed cluster, are selected for the computation of its age, mass, and extinction.}
{In the case of the M31 galaxy, the extinction range is wide and the age-extinction degeneracy is strong for a fraction of its clusters. Because of a narrower extinction range, the age-extinction degeneracy is weaker for the M33 clusters. By using artificial cluster sample, we show that age-extinction degeneracy can be reduced significantly if the range of intrinsic extinction within the host galaxy is narrow.}
{}

\maketitle

\section{Introduction}
Star clusters are important objects for understanding the formation and evolution of their host galaxies. It is considered that most of star formation is clustered \citep{Lada2003}, therefore knowledge of the physical parameters of a cluster population (e.g., age, mass, chemical composition, and extinction) is essential for constraining the star formation history of the galaxy.

The commonly used method for deriving the physical parameters of unresolved star clusters is based on comparing of their integrated broad-band photometry colors to the colors of simple stellar population (SSP) models; see, e.g., \cite{Anders2004} and \cite{Bridzius2008}. However, this method is strongly biased by the presence of two major problems:
\begin{itemize}
\item degeneracy between various physical parameters of star clusters, see, e.g., \cite{Worthey1994}, \cite{Bridzius2008}. For example, a young cluster possessing high extinction can have colors similar to an older object without extinction -- an age-extinction degeneracy;
\item stochasticity, which is due to the random presence of a few bright stars, which dominate the integrated photometry of unresolved clusters; see, e.g., \cite{Santos1997}, \cite{Deveikis2008}, \cite{MaizApellaniz2009}. Young clusters can have supergiant stars that significantly redden their integrated colors. By using the SSP method, a much older age would be determined for these clusters.
\end{itemize}

Although \cite{Cervino2006} have attempted to describe the problem of stochasticity analytically, the nowadays preferred approach (e.g., \cite{Popescu2009, Popescu2010}; \cite{Fouesneau2010,Fouesneau2012}; \cite{Asad2012}) is to use a Monte-Carlo sampling of stellar IMF to model integrated colors of clusters and build an extensive grid of models, to cover all possible age, mass and extinction ranges. Physical parameters of star clusters are then derived by comparing observations to the grid of models.

Recently \cite{Asad2012} have derived the age and extinction of star clusters in the Large Magellanic Cloud (LMC) using the method of \citet{Popescu2010} that takes stochasticity into account, and compared results to previous derivations based on an isochrone fit to the resolved color-magnitude diagrams (CMDs). They managed to constrain the age of clusters similar to the values found by the isochrone fit only when previously known extinctions of individual clusters were used.

In this paper, a method of deriving physical parameters is developed and applied to star cluster samples (integrated multi-band photometry) in two Local Group galaxies: 1) M31, using a catalog by \cite{Vansevicius2009}, who derived cluster parameters using SSP method, and 2) M33, using objects common to catalogs of \cite{Ma2012} for photometry and \cite{SanRoman2009}, who observed in resolved conditions with the {\it Hubble Space Telescope} ({\it HST}) and estimated age, mass, and extinction based on an isochrone fit to cluster CMDs.

Using this data and artificial cluster simulations, we demonstrate that when the intrinsic range of extinction within the host galaxy is rather narrow, it is not necessary to know the exact value of the extinction for individual clusters to derive their physical parameters reliably.

The paper is organized into the following sections. Section \ref{sec:method} introduces our method of deriving physical parameters of clusters when stochasticity is taken into account, \S\,\ref{sec:grid} describes a grid of simulated cluster models, \S\,\ref{sec:artificial} presents tests of the method on artificial cluster samples, and \S\,\ref{sec:application} applies the method to the M31 and M33 star clusters.

\section{Method of deriving age, mass, and extinction of star cluster}
\label{sec:method}
There are presently two main methods used to derive physical parameters of star clusters based on a 3D grid (age, mass and extinction) of models, which take stochasticity effects into account. The first is a fast $\chi^{2}$ minimization approach used by, e.g., \cite{Popescu2010} and \cite{Beerman2012}, the second, a more accurate but much slower approach, which builds probability maps in the age-mass-extinction space by exploring all the nodes of the grid and selects the most probable solution    \citep[see e.g., ][]{Fouesneau2010}.

\begin{figure}
\centering
\includegraphics[width=88mm]{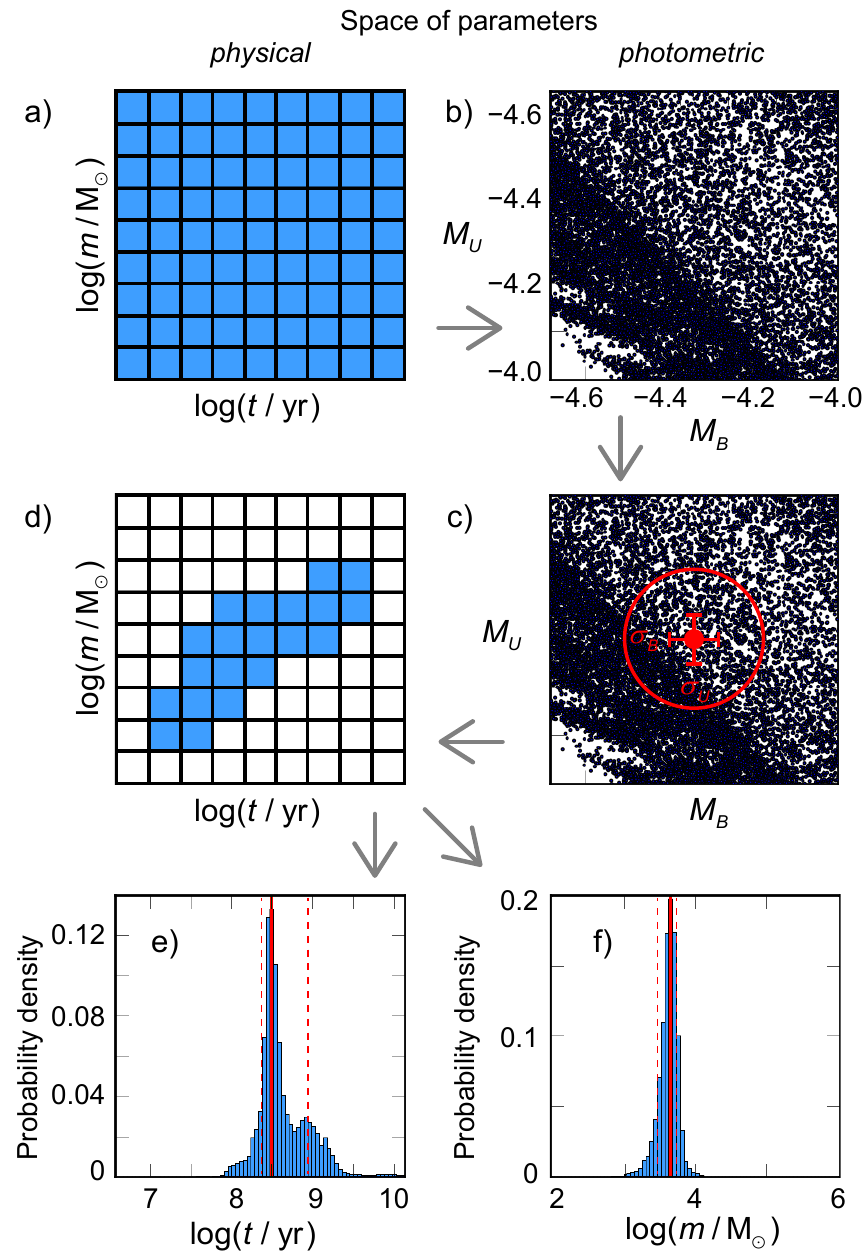}
\caption{\small Scheme of the method for deriving physical parameters of star cluster. Panel (a) shows the grid of models (only age and mass are displayed; same for extinction). These models are represented in the photometric space (only $M_{U}$ and $M_{B}$ are shown) displayed in panel (b; only 100 models per node without extinction are shown). In panel (c) all the models situated within 3--$\sigma$ (circle) from the observed absolute magnitudes are selected. Panel (d) shows the nodes associated with these models. The 1D age and mass distributions of the selected models are displayed as normalized histograms in panels (e) and (f), used to determine most probable values, indicated by solid vertical lines; confidence intervals are indicated by dashed lines.}
\label{fig1}
\end{figure}

The method presented here also explores parameter probability maps; however, by restricting the analysis to the models located in the vicinity of the observed absolute magnitudes (the distance to the object has to be known), we significantly reduce the computation time. The scheme of the method is sketched in Fig.\,\ref{fig1}.

A 3D grid of cluster models is built for every value of the three physical parameters $t,\,m,\,E(B-V)$\,\footnote{We refer to extinction as $E(B-V)$ hereafter.}; for simplicity Fig.\,\ref{fig1}\,(a) shows only a grid for age and mass. For the description of the grid, see \S\,\ref{sec:grid}. Each node of the grid contains 1\,000 models of the same age, mass, and extinction. They populate the photometric parameter space (absolute $UBVRI$ magnitudes). Figure \ref{fig1}\,(b) shows $M_{U}$ and $M_{B}$ with only 100 models per node without extinction, and the used model grid is much more continuous in photometric parameter space.

When the observations are considered in Fig.\,\ref{fig1}\,(c), along with their error bars ($\sigma$; hereafter we use $\sigma = 0.05$ mag for all the passbands of artificial and real cluster samples studied in this paper), which in general can be different for every magnitude, all the models situated within 3--$\sigma$ from the observed magnitudes are selected. Fig.\,\ref{fig1}\,(d) shows the nodes to which the selected models are associated. Other nodes do not play any role in the derivation of parameters, therefore the speed of the algorithm is increased significantly. Finally, the distributions of age and mass, displayed in Figs. \ref{fig1}\,(e, f), are derived from the selected models, as for the extinction, which is not shown in the figure.

For the selected models (Fig.\,\ref{fig1}\,(c) circle) we apply weights as follows: for the models located within 1--$\sigma$ from the observed magnitude a weight of 0.68 is assigned, for the ones between 1--$\sigma$ and 2--$\sigma$ -- 0.28, and for the ones between 2--$\sigma$ and 3--$\sigma$ -- 0.04. The probability density distributions displayed in Figs. \ref{fig1}\,(e, f) are derived by normalizing the total area of each histogram to 1. The solution is taken as the maximum of these 1D distributions. We compute confidence intervals (error bars) by excluding the first and the last 16\% of the area in histograms, following the method of ``central interval'' presented in \S\,2.5.1 of \cite{Andrae2010}.

\section{The age-mass-extinction grid of models}
\label{sec:grid}
To derive physical parameters of star clusters with the method described in \S\,\ref{sec:method}, a large age-mass-extinction grid of models is computed, by applying the algorithm described by \cite{Deveikis2008}. The stellar masses are generated randomly sampling the IMF \citep[corrected for binaries; ][]{Kroupa2001} and their luminosities are derived from stellar isochrone of the selected age and metallicity of the cluster model. The process is continued until the total mass of generated stars reaches the mass of the cluster model. Then, taking the distance to the cluster into account, the magnitudes are computed using the Johnson-Cousins $UBVRI$ photometric system \citep{MaizApellaniz2006}.

For stellar models, we took the PADOVA isochrones\footnote{PADOVA isochrones from ``CMD 2.4'': http://stev.oapd.inaf.it/cmd} from \cite{Marigo2008}, with corrections by \cite{Girardi2010} for the TP-AGB phases. The model grid for a single metallicity $Z=0.008$ contains the following nodes: ages from $\log(t/{\rm yr})$ = 6.6 to 10.1 in steps of 0.05, masses from $\log(m/\msol)$ = 2 to 5 in steps of 0.05. This gives 71 values of age and 61 values of mass, and the grid consists of 1\,000 models per node, i.e. $\sim$ $4\times10^{6}$ stochastic models. To limit the number of models to store in computer's memory, extinction is computed when the observed cluster is compared with the grid of models. It ranges from $E(B-V)$ = 0 to 1 in steps of 0.02, therefore 51 values for the extinction.

To accelerate computation of integrated magnitudes of stochastic star cluster models, we defined a threshold in the isochrone, under which the total luminosity of fainter stars is computed by integration of the stellar luminosity function along the isochrone. Above the threshold, which is defined to select 20\% of the most massive stars, the contribution of the high-mass stars is modeled by the algorithm of \cite{Deveikis2008}. The models built by this improved procedure share the same properties as models built by simulating all stars of the cluster, but require much less computation with a speed gain of a factor $\sim$10.

\section{Test of the method on an artificial cluster sample}
\label{sec:artificial}
\subsection{Degeneracies in an artificial cluster sample}
We simulated artificial star cluster samples with known age, mass, and extinction and used them as input clusters to evaluate the ability of our method to derive physical parameters. The artificial cluster samples consist of 10\,000 clusters with a random age in the $\log(t/{\rm yr})$ range of [6.6, 10.1]. To simulate the mass of input clusters, we used a power-law cluster mass function with index $-2$ in the range of $\log(m/\msol)$ = [2.7, 4.3], so as to have more low-mass clusters in the sample. We have two artificial samples: one without extinction and the other with $E(B-V)$ in the range of [0.0, 1.0] using the Milky Way standard extinction law from \cite{Cardelli1989}.

\begin{figure}
\centering
\includegraphics[width=88mm]{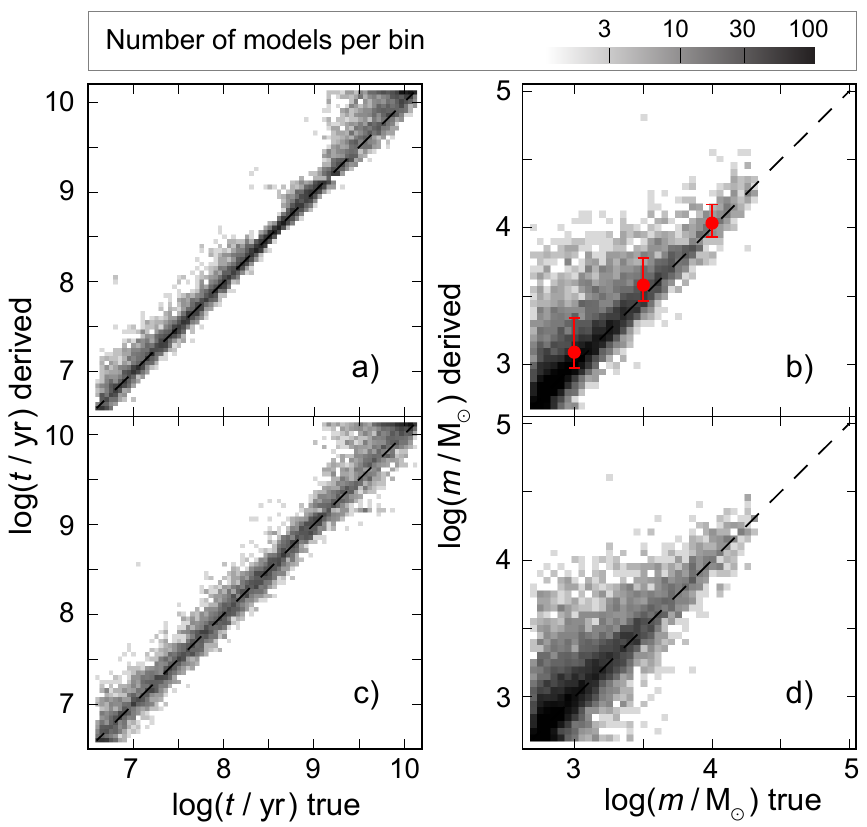}
\caption{\small Derived parameters of 10\,000 artificial clusters without extinction. Panels (a) and (b) show age and mass for a sample without photometric errors; (c) and (d) for a sample with Gaussian photometric errors of 0.05 mag. Panel (b) also shows the means and standard deviations (circles with error bars) of the derived mass distributions for clusters with true mass $\log(m/\msol)$ = 3.0, 3.5 and 4.0. The density scale is logarithmic.}
\label{fig2}
\end{figure}

Figure \ref{fig2} displays the results for artificial cluster sample built without extinction. Panels (a) and (b) show results of the age and mass of the artificial cluster sample without introducing photometric observation errors. Panels (c) and (d) show the case with Gaussian photometric errors of 0.05 mag randomly added to each magnitude of the sample clusters. The introduction of photometric errors results in broadening around the one-to-one line.

In Fig.\,\ref{fig2}\,(b), the asymmetry observed in the mass derivation slightly favors high masses. This is because that in the grid of star cluster models, the nodes of models with higher mass have magnitudes that are less dispersed than the nodes with lower mass models, as a consequence of stochasticity \cite[see e.g., ][]{Deveikis2008}. Thus, when the models of two nodes of different masses are located in the $UBVRI$ ``sphere'' around the observation shown in Fig.\,\ref{fig1}\,(c), the more massive node, with less dispersed models, will dominate. To balance out the effect, a cluster mass function could be used to decrease the importance of the nodes of massive cluster models. Although Fig.\,\ref{fig2}\,(b) shows asymmetry, the means and standard deviations given at $\log(m/\msol)$ = 3.0, 3.5, and 4.0 indicate that this phenomenon is slight.

\begin{figure*}
\centering
\includegraphics[width=180mm]{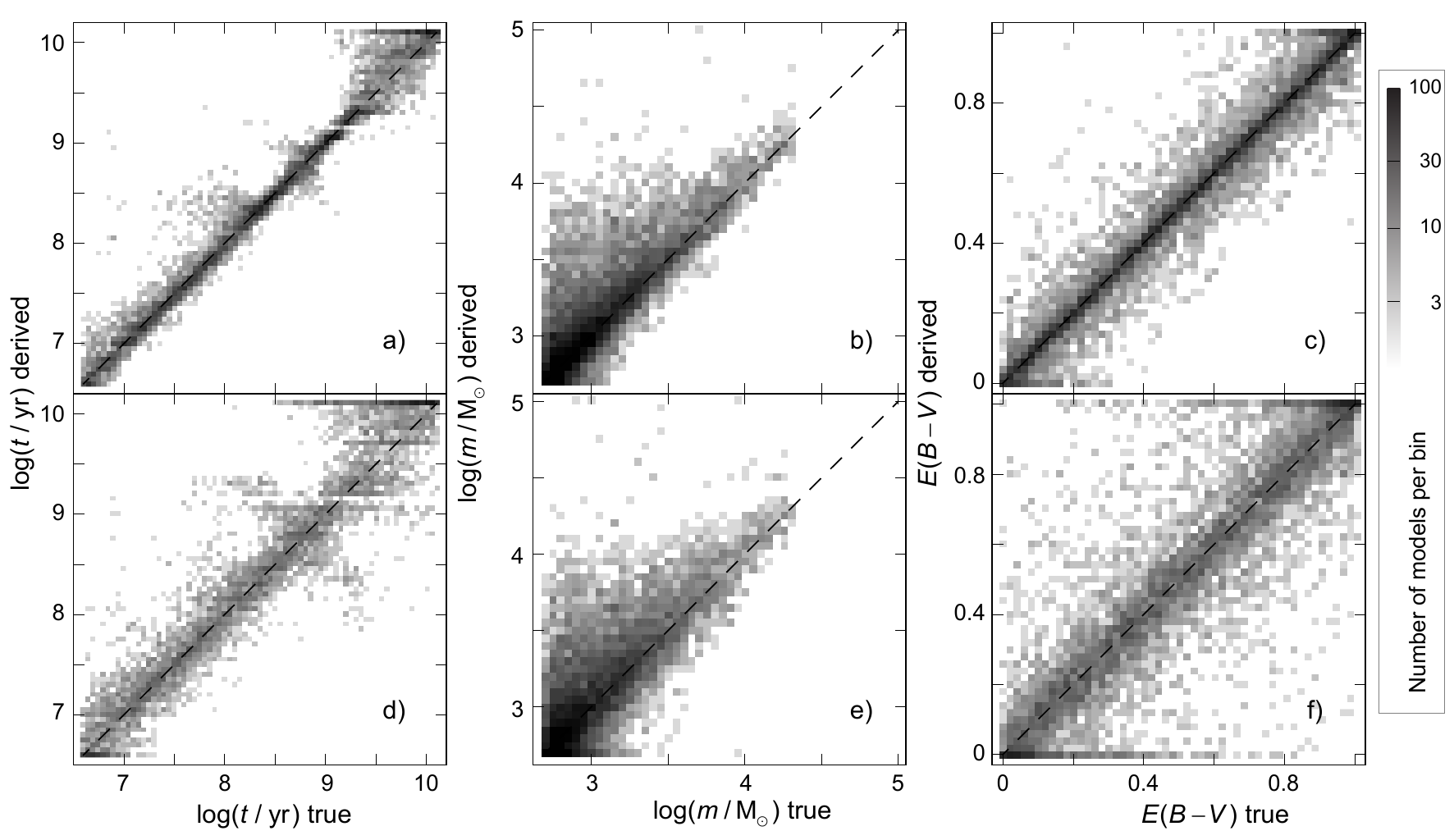}
\caption{\small Derived parameters of 10\,000 artificial clusters with extinction randomly chosen in the range of $E(B-V)$ = [0, 1]. Panels (a), (b) and (c) show age, mass, and extinction for a sample without photometric errors; (d), (e), and (f) for a sample with Gaussian photometric errors of 0.05 mag.}
\label{fig3}
\end{figure*}

Figure \ref{fig3} gives the results of artificial clusters in the case of extinction. Panels (a), (b), and (c) show the age, mass and extinction derived when there is no photometric errors, and panels (d), (e), and (f) when there are Gaussian photometric errors of 0.05 mag included. In Fig.\,\ref{fig3}\,(d), in the case of photometric errors and extinction, we observe that a broadening around the one-to-one relation increases, especially for $\log(t/{\rm yr}) \gtrsim 8$, which is associated with broadening in the extinction (panel f), which is a sign of the age-extinction degeneracy. It creates two streaks above and below the one-to-one relation in the range of $8 \lesssim \log(t/{\rm yr}) \lesssim 9.5$ that were already perceptible in the case without photometric errors in Fig.\,\ref{fig3}\,(a). However, including of photometric errors does not significantly affect the derivation of mass (panels b and e). We note that a gap in derived ages at $\log(t/{\rm yr}) = 9.15$ is a feature of isochrone due to the increase in the production rate of AGB stars, which was discussed in \cite{Girardi1998}.

These degeneracy streaks have first been reported by \cite{Fouesneau2010}. The degeneracy streak above the one-to-one line seen in Fig.\,\ref{fig3}\,(d) concerns clusters that are young and that possess intrinsically high extinction, but are derived by the method as older and having lower extinction. Conversely, the streak of clusters below the one-to-one relation involves objects that are derived as older and that have lower extinction than they do in reality. These features are important to keep in mind when deriving the physical parameters of unresolved star clusters.

The treatment of the metallicity effects on the derivation of age, mass, and extinction of star clusters is the subject of a forthcoming paper. To illustrate the effect, we derived the parameters of a cluster sample of $Z = 0.008$ metallicity, successively with a model grid of much lower metallicity, $Z = 0.00013$, and another one with much higher metallicity, $Z = 0.03$. In the first case, the ages are systematically overestimated by $\sim$0.5 dex, the upper streak is more developed, and the lower one disappears. The masses are also overestimated by $\sim$0.5 dex, and there is a preference for overestimated extinction. In the second case, the ages are only slightly ($\sim$0.1 dex) underestimated. The upper streak decreases without vanishing, and the lower one becomes more populated. The masses are underestimated of $\sim$0.1 dex, and the extinction is almost unaffected.

\subsection{Is it possible to reduce the age-extinction degeneracy?}
The upper and lower streaks in Fig.\,\ref{fig3}\,(d) suggest that if a wide extinction range is allowed in a simulated sample, then there are possibilities that a cluster mimics an older one with lower extinction, or inversely a younger one with higher extinction. If the true extinction range of the cluster population is narrow, then we could restrict the search for the extinction within a narrow range in the model grid, resulting in decrease of age-extinction degeneracy.

\begin{figure*}
\centering
\includegraphics[width=180mm]{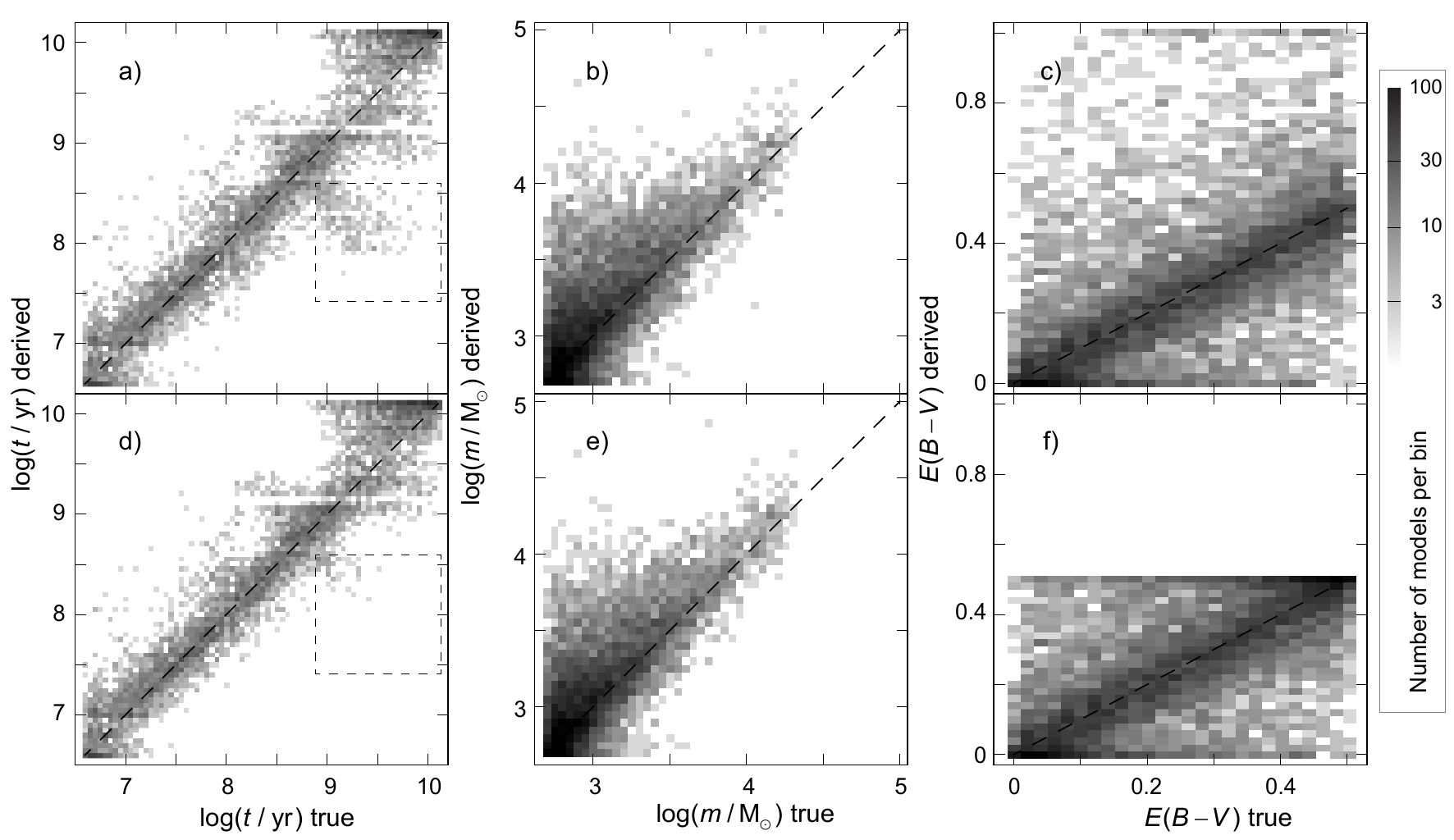}
\caption{\small Derived parameters of 10\,000 artificial clusters with true extinction randomly chosen in the range of $E(B-V)$ = [0, 0.5] with Gaussian photometric errors of 0.05 mag. Panels (a), (b), and (c) show age, mass and extinction when a sample is studied using a grid of models with an allowed extinction in the range of [0, 1]; (d), (e), and (f) show results for the same cluster sample studied with a narrower allowed extinction in the range of [0, 0.5]. Panels (c) and (f) are denser than in Fig.\,\ref{fig3}, because the same number of clusters are located on a smaller range of extinction. A square indicated in panels (a) and (d) is used to quantify the degeneracy effect.}
\label{fig4}
\end{figure*}

In Fig.\,\ref{fig4} we show results of the tests for a sample of 10\,000 artificial clusters with true extinction from a range of $E(B-V)$ = [0, 0.5] and with Gaussian photometric errors of 0.05 mag randomly added to each magnitude of the sample clusters. This cluster sample was studied twice, with different allowed extinction ranges in the model grid.

In the first test, the extinction of the model grid was allowed to vary in a wide range, $E(B-V)$ = [0, 1], shown in Fig.\,\ref{fig4}\,(a, b, c). If a cluster has a true extinction of 0.5, then the maximum underestimation of its extinction can be from 0 to 0.5. But if a cluster has true extinction of 0, the maximum overestimation of its extinction could range from 0 to 1. This explains why the lower streak (i.e. clusters with overestimated extinction) is more extended than the upper one in panel (a).

In the second test, the allowed extinction range of the model grid was reduced to a range of [0, 0.5]; Fig.\,\ref{fig4}\,(d, e, f). The constraint on the extinction range resulted in a reduction of the lower degeneracy streak, seen in Fig.\,\ref{fig4}\,(d), which is less developed than in Fig.\,\ref{fig4}\,(a). From the comparison of Figs.\,\ref{fig4}\,(b) and (e), we see that the mass is less affected by the degeneracy. We note that only the lower streak is modified, since only the higher limit of the allowed extinction range was changed from 1.0 to 0.5. The upper streak is not modified, because the lower limit of the allowed extinction range was not changed.

To quantify the reduction of the lower degeneracy streak due to reduction of the extinction range, all the models situated in the square shown in Figs.\,\ref{fig4}\,(a) and (d) were counted. There are $\sim$480 clusters in that region for the wide extinction range (panel a) and only $\sim$110 for the low extinction range (panel d). The reduction of the extinction range from $E(B-V)$ = [0, 1] to [0, 0.5] thus decreases more than four times the strength of the degeneracy streak.

We conclude that to reduce the degeneracy streaks seen in Fig.\,\ref{fig3}\,(d), we should make a reasonable assumption on the extent of the possible extinction range within the galaxy hosting the studied cluster population.

\section{Application of the method to real star clusters}
\label{sec:application}
\subsection{The M31 galaxy star cluster sample}
A star cluster catalog of 285 objects located in the south west field of the M31 galaxy was compiled by \cite{Narbutis2008} using the deep $BVRI$ and $H_{\alpha}$ photometry from the Subaru telescope, as well as multiband maps based on {\it HST}, {\it Galex}, {\it Spitzer}, and 2MASS imaging. The $UBVRI$ photometry was derived using the Local Group Galaxy Survey data from \cite{Massey2006}. The magnitude limit of the cluster sample was set to the $V = 20.5$ mag.

\begin{figure*}
\centering
\includegraphics[width=180mm]{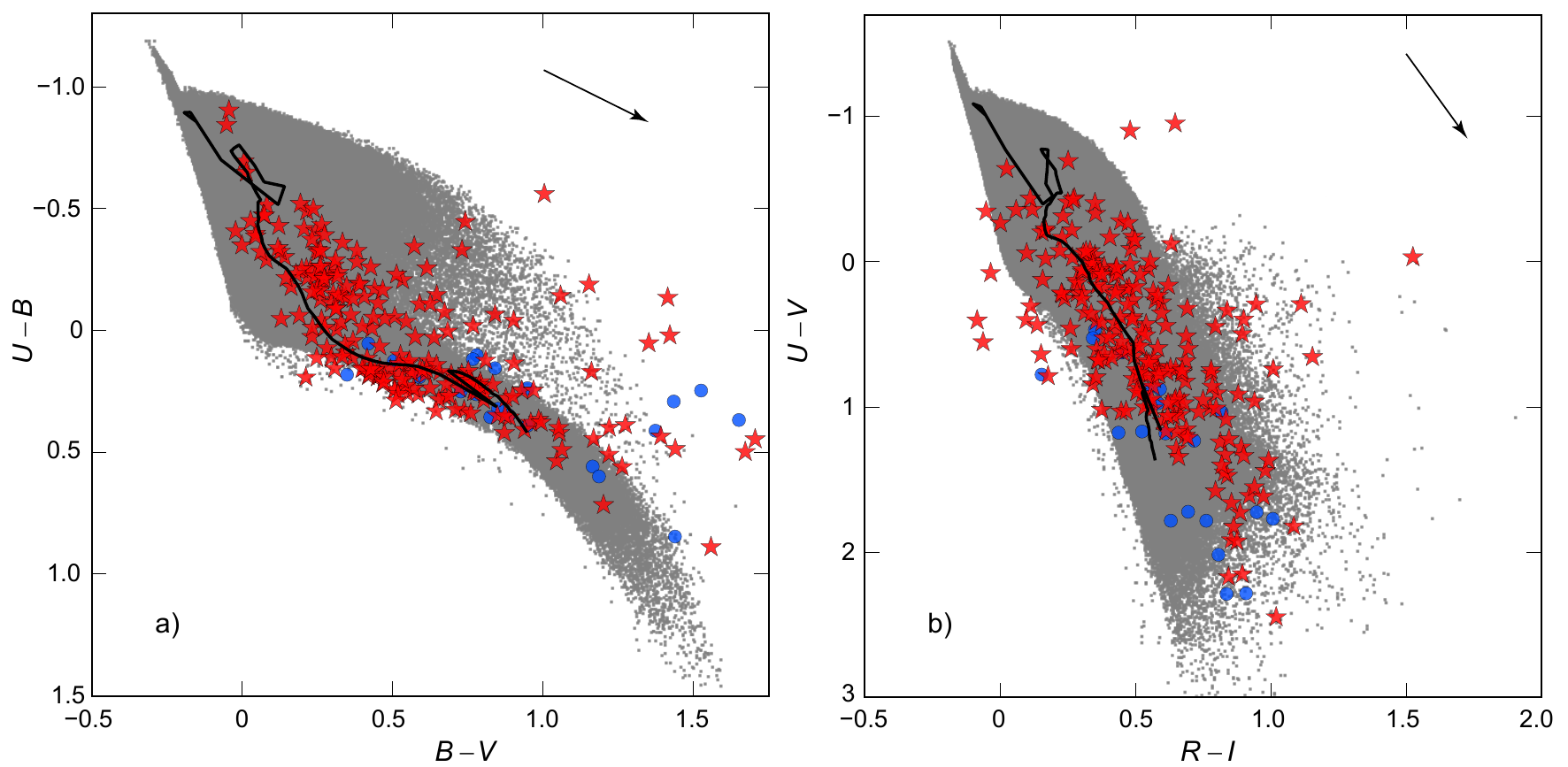}
\caption{\small M31 galaxy star cluster sample of 238 objects from \cite{Vansevicius2009} (circles and stars) and the studied sample of 216 objects (stars) having galactocentric distance over 7 kpc. Panels show: (a) $U-B$ vs $B-V$ and (b) $U-V$ vs $R-I$ diagrams. The model grid (without extinction) used to derive physical parameters of clusters is displayed in the background with indicated reddening vectors following the Galactic extinction law. The continuous line traces PADOVA SSP of $Z=0.008$.}
\label{fig5}
\end{figure*}

\cite{Vansevicius2009} selected 238 clusters from that sample, excluding the ones with strong $H_{\alpha}$ emission, and compared their multiband colors to PEGASE \citep{Fioc1997} SSP models to derive their age, mass, metallicity, and extinction. {\it Spitzer} data was used to constrain the maximum extinction for each cluster. They reported $\sim$30 classical globular clusters with low metallicity older than 3 Gyr. The remaining $\sim$210 younger clusters were classified as objects belonging to the disk, with average metallicity $Z=0.008$. Figure \ref{fig5} shows the $U-B$ vs $B-V$ and $U-V$ vs $R-I$ diagrams of the 238 star clusters from \cite{Vansevicius2009}, compared to the grid of star cluster models built in \S\,\ref{sec:grid}.

Since our grid of models has single metallicity, $Z=0.008$, we attempted to exclude more metal-rich clusters from \cite{Vansevicius2009} sample by only selecting objects with galactocentric distance over 7 kpc. This subsample consists of 216 clusters and is displayed in Fig.\,\ref{fig5}. It is studied with our method described in \S\,\ref{sec:method}, using the Milky Way standard extinction law \citep{Cardelli1989} and distance modulus to M31 of $(m-M)_{0} = 24.47$ derived by \cite{McConnachie2005}.

\begin{figure*}
\centering
\includegraphics[width=180mm]{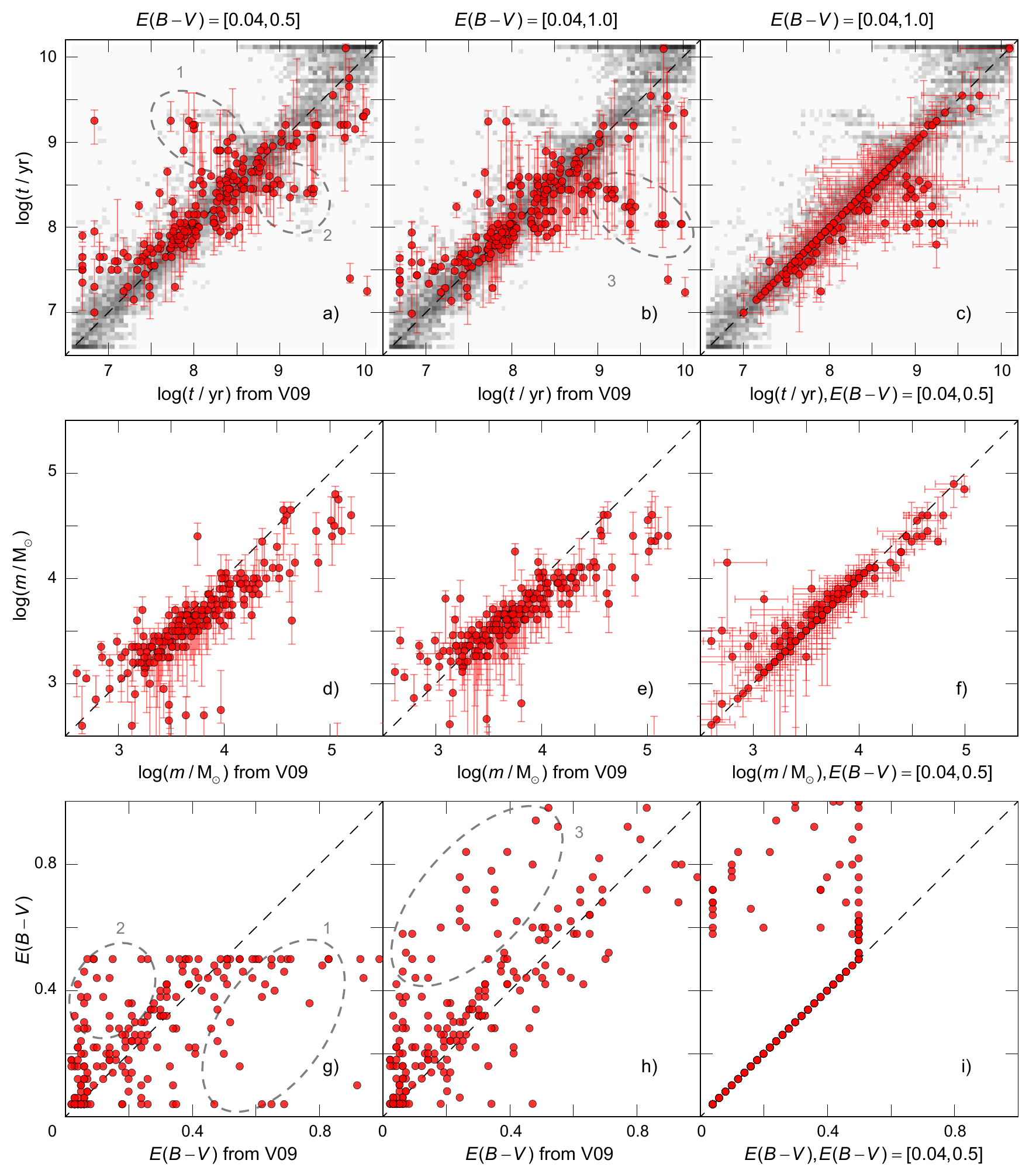}
\caption{\small M31: age, mass, and extinction of 211 star clusters from \cite[][as V09]{Vansevicius2009}. Panels (a, d, g) show results when a narrow extinction range of $E(B-V)$ = [0.04, 0.5] is allowed in the model grid, compared to V09 values. Panels (b, e, h) show results for a wider allowed extinction range [0.04, 1.0], compared to V09 values. Panels (c, f, i) compare results obtained in a wide extinction range vs the ones obtained in a narrow extinction range. In panels (a, b, c), a density map of Fig.\,\ref{fig3}\,(d) is reproduced to show regions of age-extinction degeneracy. Error bars are computed as described in \S\,2. Six clusters from V09, which have $E(B-V) > 1$, are not shown in panels (g) and (h). Dashed ellipses, numbered ``1'', ``2'', and ``3'', describe majority of associated clusters in the age and extinction panels.}
\label{fig6}
\end{figure*}

Figure \ref{fig6} presents the age (panels a, b, c), mass (panels d, e, f) and extinction (g, h, i) of 211 clusters (from the 216 sample) derived by using our method; for five clusters, no model was found within the 3--$\sigma$ around the observed magnitudes. Clusters were studied twice, first with a narrow extinction range $E(B-V)$ = [0.04, 0.5] allowed in the model grid, and second with a wider one, [0.04, 1.0]. The first column (panels a, d, g) compares the age, mass, and extinction derived when a narrow extinction range is allowed vs the \cite{Vansevicius2009} results. The second column (panels b, e, h) compares the age, mass, and extinction derived when a wide extinction range is allowed vs the \cite{Vansevicius2009} results. The last column (panels c, f, i) compares the results obtained with a wide extinction range allowed vs the ones obtained with a narrow extinction range allowed.

In Fig.\,\ref{fig6}\,(a), when clusters are studied in a narrow allowed extinction range, $E(B-V)$ = [0.04, 0.5], the derived ages show the same features as the models studied in Fig.\,\ref{fig3}\,(d), reproduced here in the background of the panel. The degeneracy streaks develop above and below the one-to-one line, perpendicularly to it, and are marked by ellipses numbered ``1'' and ``2'' in Fig.\,\ref{fig6}\,(a). As for the models in the background, the upper degeneracy streak concerns clusters with overestimated age and underestimated extinction, shown by ellipse ``1'' in Fig.\,\ref{fig6}\,(g). In contrast, for the lower degeneracy streak (ellipse ``2''), the age is underestimated and extinction is overestimated. Since the upper streak (ellipse ``1'') is more developed than the lower one (``2''), we interpret that as a result of too narrow an extinction range in the model grid allowed $E(B-V)$ = [0.04, 0.5].

In Fig.\,\ref{fig6}\,(b), when clusters are studied in a wider model extinction range, $E(B-V)$ = [0.04, 1], the upper degeneracy streak is less populated, and the lower one extends, meaning that some of the clusters have overestimated extinction (ellipse ``3''), also shown in panel (h).

If the intrinsic range of extinction for a cluster sample is wide, as in the galaxy M31, then a narrow extinction range of models produces extended an upper streak (ellipse ``1'') and a smaller lower streak (ellipse ``2''), shown in Fig.\,\ref{fig6}\,(a). In panel (b), when the allowed extinction range of models is wide, the upper streak retracts and the lower streak develops. We conclude that in a galaxy with wide extinction range, it is not possible to derive parameters for the clusters affected by age-extinction degeneracy and additional constraints for the extinction are needed; e.g., \cite{Vansevicius2009} used a {\it Spitzer} emission map, to trace the dust lanes of M31 and reduce the age-extinction degeneracy.

Figs. \ref{fig6}\,(d, e) show that for high-mass clusters we obtain lower masses than given by \cite{Vansevicius2009}. Inspecting the metallicity values provided by \cite{Vansevicius2009}, we note that high-mass clusters have metallicities lower than the $Z = 0.008$ used in our model grid. This could be a sign of the metallicity effect on the derivation of physical parameters, and will be investigated in the forthcoming paper.

\subsection{The M33 galaxy star cluster sample}
\cite{SanRoman2009} reports observing of 161 clusters using the {\it HST}, allowing them to evaluate their ages and extinctions by PADOVA \cite{Girardi2002} isochrone fit to their resolved CMDs. Recently, \cite{Ma2012} has used 392 clusters from the compiled catalog of \cite{Sarajedini2007} and images from \cite{Massey2006} to provide $UBVRI$ broad-band integrated photometry data.

There are 40 clusters common to both the \cite{SanRoman2009} and the \cite{Ma2012} catalogs, making them interesting to compare the parameters derived from the resolved method (isochrone fit to CMDs by \cite{SanRoman2009}) and stochastic method, which was applied here to the integrated photometry data of \cite{Ma2012}.

\begin{figure*}
\centering
\includegraphics[width=180mm]{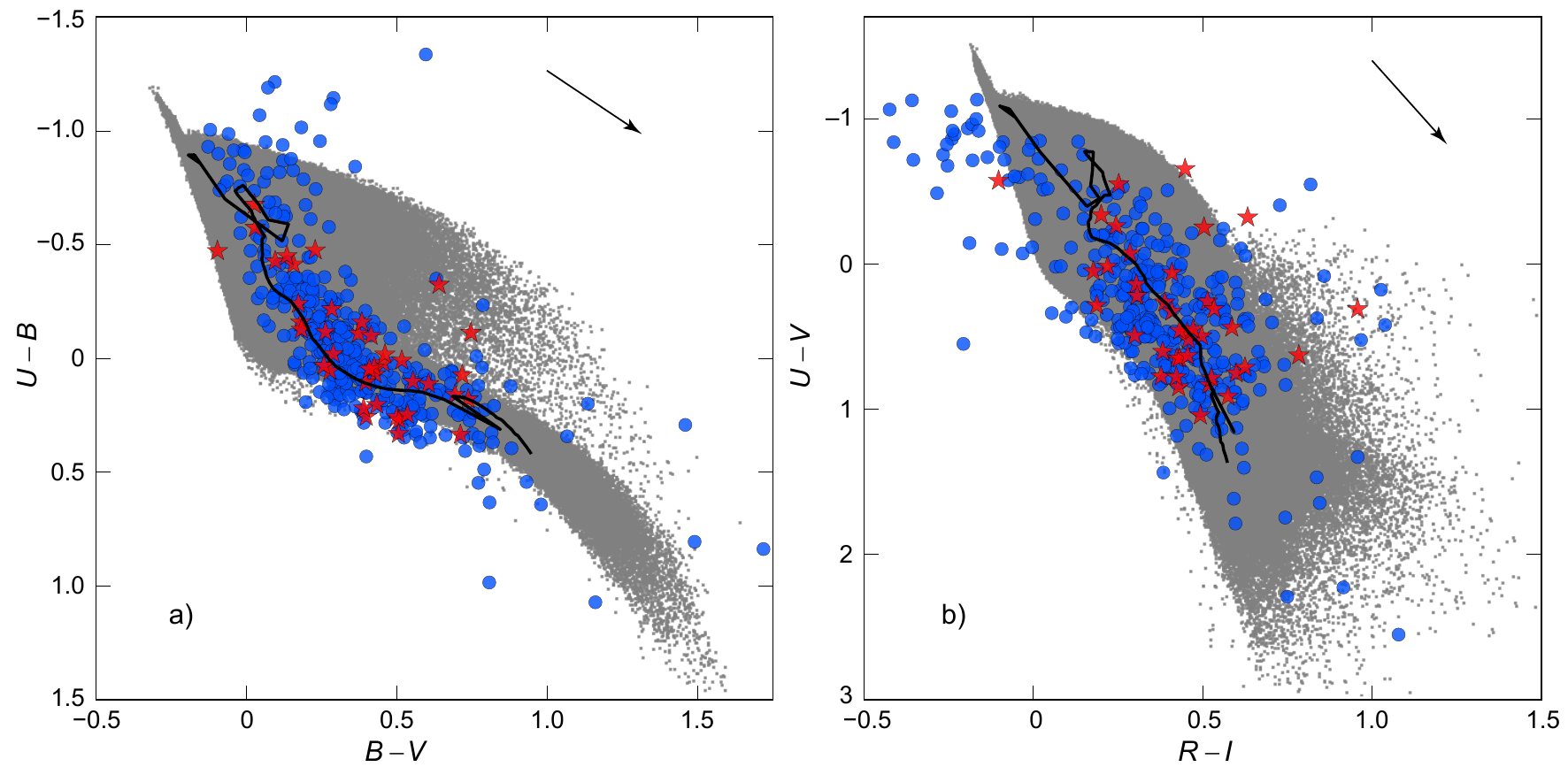}
\caption{\small M33 galaxy star cluster sample of 161 objects from catalog of \cite{SanRoman2010} (circles and stars) and the studied sample of 40 objects (stars) common to \cite{Ma2012} and \cite{SanRoman2009}. Panels show (a) $U-B$ vs $B-V$, and (b) $U-V$ vs $R-I$ diagrams. The model grid (without extinction) used to derive physical parameters of clusters is displayed in the background with indicated reddening vectors following the LMC extinction law. The continuous line traces PADOVA SSP of $Z=0.008$. The $B$ band magnitudes of clusters were shifted by 0.1 mag.}
\label{fig7}
\end{figure*}

Figure \ref{fig7}\,(a) presents the $U-B$ vs $B-V$ diagram of the 392 clusters from the \cite{Ma2012} catalog compared to the grid of cluster models. The selected 40 star clusters are also indicated. To account for calibration of the $B$ band, as discussed by \cite{Ma2012}, where he compared his photometry to previous works by \cite{Sarajedini2007}, \cite{Park2007}, and \cite{SanRoman2009}, we shifted the $B$ band to brighter magnitudes by 0.1 mag for all clusters. We note that the parameter derivation results do not change when the $B$ band is not taken into account.

Since the metallicity content of the M33 galaxy is similar to the one of the LMC \citep{Bresolin2010}, we assume that the interstellar extinction law derived by \cite{Gordon2003} for the LMC can be applied to the M33 cluster sample. As in \cite{SanRoman2009}, we adopted the M33 distance modulus of $(m-M)_{0} = 24.54$ \citep{McConnachie2005}.

For the M33, \cite{U2009} derived the extinction of 22 supergiants, which resulted in an extinction distribution centered on $E(B-V)$ = 0.1. \cite{U2009} also used the data of \cite{Rosolowsky2008} to derive $E(B-V)$ values for 58 HII regions, and show in their Fig.\,9 that extinction can be expected to be $E(B-V)$ $\lesssim$ 0.3 for those regions (except for 3 objects), with an average $E(B-V)$ $\sim$ 0.11. A foreground Galactic line-of-sight extinction in the direction of M33 of $E(B-V)$ = 0.04 is estimated from \cite{Schlegel1998} extinction maps.

We studied the M33 clusters using two different allowed extinction ranges for the model grid. The first one is narrow, extending between the foreground Galactic line-of-sight extinction in the direction of M33 up to the higher limit given by \cite{U2009}, i.e, $E(B-V)$ = [0.04, 0.30]. The second one is a wider extinction range, $E(B-V)$ = [0.04, 1.0], such as the one used for the M31 case.

\begin{figure*}
\centering
\includegraphics[width=180mm]{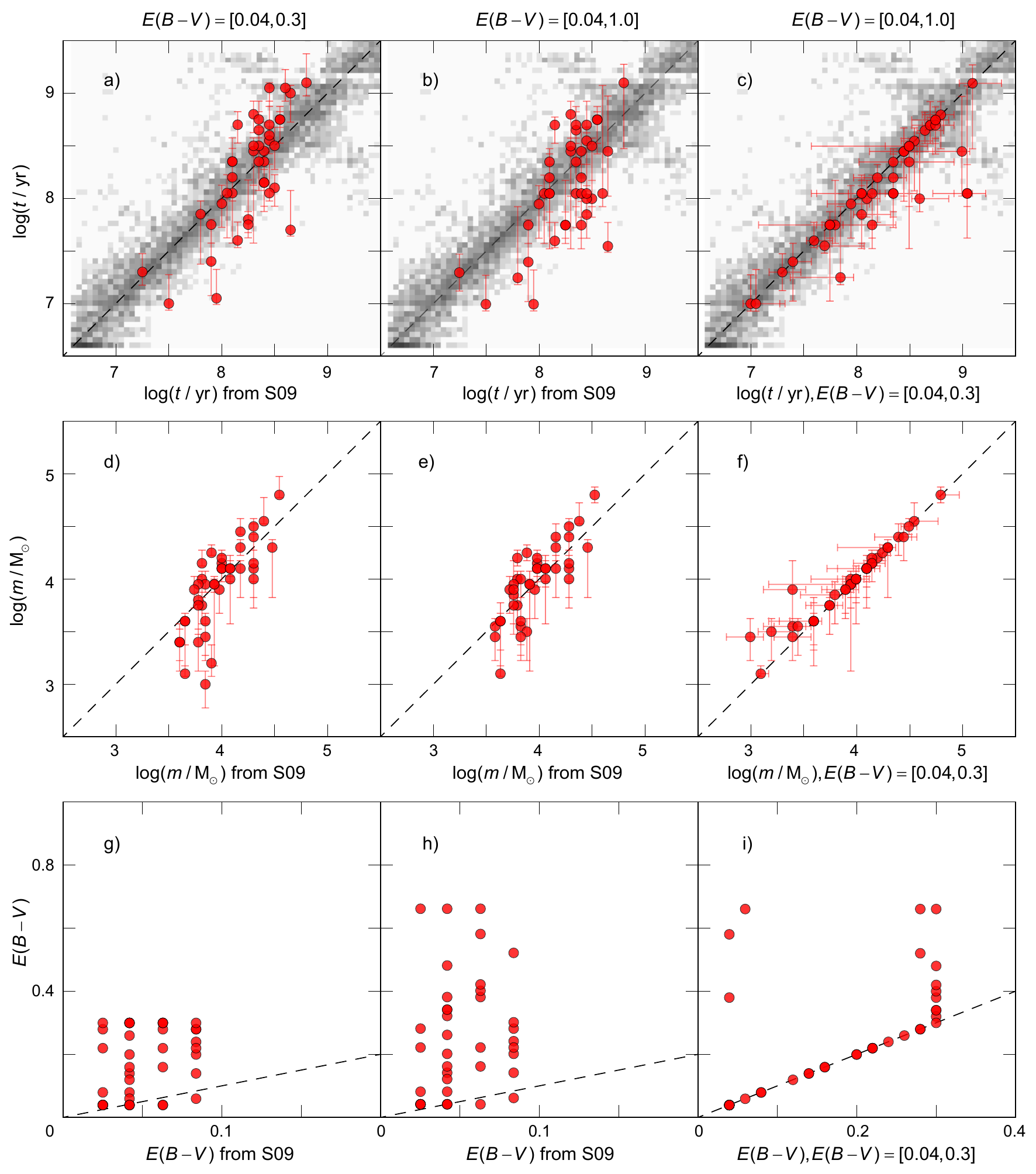}
\caption{\small M33: age, mass and extinction of 40 star clusters common to catalogs of \cite{Ma2012} and \cite[][as S09]{SanRoman2009}. Panels (a, d, g) show results when a narrow extinction range of $E(B-V)$ = [0.04, 0.3] is allowed in the model grid, compared to S09 values. Panels (b, e, h) show results for a wider allowed extinction range of $E(B-V)$ = [0.04, 1.0], compared to S09 values. Panels (c, f, i) compare results obtained in a wide extinction range vs the ones obtained in a narrow extinction range. In panels (a, b, c), a density map of Fig.\,\ref{fig3}\,(d) is reproduced to show regions of age-extinction degeneracy. Error-bars are computed as described in \S\,2.}
\label{fig8}
\end{figure*}

Figure \ref{fig8} presents the age (panels a, b, c), mass (panels d, e, f), and extinction (g, h, i) of the 40 clusters derived by using our method. The results obtained in the narrow extinction range (panels a, d, g) and the wide extinction range (panels b, e, h) are compared to the ones obtained by \cite{SanRoman2009} using isochrone fit to CMDs. Panels (c, f, i) compare the results we obtained with the wide extinction range allowed vs the narrow extinction range allowed. It shows that the difference is relatively small between the two kinds of solution, showing that the age-degeneracy is playing a minor role in this M33 cluster sample.

The derived mass does not differ significantly, as seen in Fig.\,\ref{fig8}\,(f). Panels (g, h, i) reveal that only one fourth of the clusters are affected by an increase in extinction when the allowed extinction range is widened. Only some of them have a much higher derived extinction when the wide allowed extinction range is used. These few clusters affected by the age-extinction degeneracy can be seen in panel (c), appearing younger when a wide extinction range is allowed than when the range is narrow, as already seen in M31 cluster sample in Fig.\,\ref{fig6}\,(c).

\section{Conclusions}
We presented a method that aims to derive the age, mass, and extinction of unresolved star clusters by using broad-band photometry and taking the stochastic sampling of stellar masses in clusters into account. We investigated the behavior of the method on a sample of artificial star clusters, in order to trace the different degeneracies between parameters, for different choices of photometric errors and extinction.

These tests allowed us to quantify the age-extinction degeneracy. We demonstrated that for the intrinsically narrow extinction range of the star cluster sample, the age-extinction degeneracy can be resolved even in cases where individual exact extinction values are not known for each cluster.

The age-extinction degeneracy has been observed in the real star cluster sample of \cite{Vansevicius2009} for M31 and to a lesser extent in the one composed of clusters common to the \cite{Ma2012} and \cite{SanRoman2009} catalogs for M33. The physical parameters derived by our method for different extinction ranges in each case, have been compared to the values provided in these studies, showing the impact of the age-extinction degeneracy, especially when the true extinction range of the cluster population is wide.

The M31 star cluster sample from \cite{Vansevicius2009} showed that a true extinction range in this galaxy is wide enough, so that the age-extinction degeneracy is significantly developed, making the parameters difficult to derive, especially for older ages. In such cases, it is preferable to use external constraints on the extinction of individual clusters.

The M33 star cluster sample shows little sign of age-extinction degeneracy, since there is only a small difference between solutions in narrow and wide extinction ranges. The range of allowed extinction of $E(B-V)$ = [0.04, 0.30] gives parameters consistent with the results of isochrone fit to CMDs by \cite{SanRoman2009}.

The follow-up paper will be dedicated to study of the metallicity effects to derive physical parameters of star clusters and to gain a complete understanding of degeneracies introduced by stochastic effects.

\begin{acknowledgements}
We are grateful to the anonymous referee for fruitful comments, which helped to improve the paper. This research was funded by a grant (No. MIP-102/2011) from the Research Council of Lithuania.
\end{acknowledgements}

\end{document}